\documentclass[aps,prd,twocolumn,a4paper,superscriptaddress,floatfix]{revtex4-1}
\usepackage{graphicx}
\begin{document}

\newcommand{\be}{\begin{equation}}
\newcommand{\ee}{\end{equation}}
\newcommand{\bq}{\begin{eqnarray}}
\newcommand{\eq}{\end{eqnarray}}
\newcommand{\bsq}{\begin{subequations}}
\newcommand{\esq}{\end{subequations}}
\newcommand{\bc}{\begin{center}}
\newcommand{\ec}{\end{center}}

\title{Fine-structure constant constraints on dark energy}

\author{C. J. A. P. Martins}
\email{Carlos.Martins@astro.up.pt}
\affiliation{Centro de Astrof\'{\i}sica da Universidade do Porto, Rua das Estrelas, 4150-762 Porto, Portugal}
\affiliation{Instituto de Astrof\'{\i}sica e Ci\^encias do Espa\c co, CAUP, Rua das Estrelas, 4150-762 Porto, Portugal}
\author{A. M. M. Pinho}
\email[]{Ana.Pinho@astro.up.pt}
\affiliation{Centro de Astrof\'{\i}sica da Universidade do Porto, Rua das Estrelas, 4150-762 Porto, Portugal}
\affiliation{Faculdade de Ci\^encias, Universidade do Porto, Rua do Campo Alegre 687, 4169-007 Porto, Portugal}

\date{12 March 2015}

\begin{abstract}
We use astrophysical and atomic clock tests of the stability of the fine-structure constant $\alpha$, together with Type Ia supernova and Hubble parameter data, to constrain the simplest class of dynamical dark energy models where the same degree of freedom is assumed to provide both the dark energy and (through a dimensionless coupling, $\zeta$, to the electromagnetic sector) the $\alpha$ variation. We show how current data tightly constrains a combination of $\zeta$ and the dark energy equation of state $w_0$. At the $95\%$ confidence level and marginalizing over $w_0$ we find $|\zeta|<5\times10^{-6}$, with the atomic clock tests dominating the constraints. The forthcoming generation of high-resolution ultra-stable spectrographs will enable significantly tighter constraints.
\end{abstract}
\maketitle

\section{Introduction}

The discovery of cosmic acceleration from luminosity distance measurements of Type Ia supernovas \cite{SN1,SN2} led to wide-ranging theoretical and observational efforts trying to understand and characterize it. While a cosmological constant $\Lambda$ remains the simplest explanation consistent with most observational data, the well-known fine-tuning problems associated with this solution imply that alternative scenarios should be sought and tested. The most natural alternative would involve scalar fields, an example of which is the recently discovered Higgs field \cite{ATLAS,CMS}, which would lead to dynamical dark energy.

If dynamical scalar fields are present, one naturally expects them to couple to the rest of the model, unless a yet-unknown symmetry is postulated to suppress these couplings. In particular, a coupling to the electromagnetic sector will lead to spacetime variations of the fine-structure constant $\alpha$---see \cite{uzanLR,cjmGRG} for recent reviews. There are some recent indications of such a variation \cite{Webb}, at the level of a few parts per million, which a dedicated Large Program at ESO's Very Large Telescope (VLT) is aiming to test \cite{LP1,LP3}.

Here, in the same spirit of \cite{Erminia1,Erminia2}, we discuss how astrophysical and local tests of the stability of $\alpha$ can be used as additional tests of the underlying theories, in particular if one makes the 'minimal' assumption that the same dynamical degree of freedom is responsible for the dark energy and the $\alpha$ variations. In this case observational tests of the evolution of $\alpha$ directly constrain dark energy. The future impact of these methods as a dark energy probe has recently been studied in detail in \cite{Amendola,Leite}.(Earlier, more simplistic forecasts can also be found in \cite{Old1,Old2}.) Here we show how current data already provides useful constraints.

We start by reviewing the relation between a varying $\alpha$ and dynamical energy in the case (dubbed 'Class I models' in \cite{cjmGRG}) where both are due to the same dynamical degree of freedom. We will then focus on the case of a constant dark energy equation of state, $w_0$, constraining this model with a combination of cosmological data and local and astrophysical tests of $\alpha$. This choice is in the interest of simplicity, as it minimizes the number of free parameters and is frequently used as a fiducial model for forecasts; we leave a discussion of more general models for a subsequent, more detailed publication. 

\section{Varying $\alpha$ and dark energy}

Dynamical scalar fields in an effective 4D field theory are naturally expected to couple to the rest of the theory, unless a (still unknown) symmetry suppresses this coupling \cite{Carroll,Dvali}. We assume this to be the case for the dynamical degree of freedom responsible for the dark energy. Specifically the coupling between the scalar field, $\phi$, and the electromagnetic sector stems from a gauge kinetic function $B_F(\phi)$
\begin{equation}
{\cal L}_{\phi F} = - \frac{1}{4} B_F(\phi) F_{\mu\nu}F^{\mu\nu}
\end{equation}
which one can assume to be linear,
\begin{equation}
B_F(\phi) = 1- \zeta \kappa (\phi-\phi_0)\,,
\end{equation}
(with $\kappa^2=8\pi G$): as pointed out in \cite{Dvali} the absence of such a term would require a $\phi\to-\phi$ symmetry, but such a symmetry must be broken throughout most of the cosmological evolution. Local tests of the Equivalence Principle lead to the conservative constraint on the dimensionless coupling parameter (see \cite{uzanLR} for an overview)
\begin{equation}
|\zeta_{local}|<10^{-3}\,,\label{localzeta}
\end{equation}
while in \cite{Erminia1} an independent few-percent constraint on this coupling was obtained using CMB and large-scale structure data in combination with direct measurements of the expansion rate of the universe.

With these assumptions one can explicitly relate the evolution of $\alpha$ to that of dark energy, as in \cite{Erminia1}. The evolution of $\alpha$ can be written
\begin{equation}
\frac{\Delta \alpha}{\alpha} \equiv \frac{\alpha-\alpha_0}{\alpha_0} =
\zeta \kappa (\phi-\phi_0) \,,
\end{equation}
and, since the evolution of the putative scalar field can be expressed in terms of the dark energy properties $\Omega_\phi$ and $w$ as
\begin{equation}
1+w_\phi = \frac{(\kappa\phi')^2}{3 \Omega_\phi} \,
\end{equation}
(with the prime denoting the derivative with respect to the logarithm of the scale factor), we finally obtain
\begin{equation} \label{eq:dalfa}
\frac{\Delta\alpha}{\alpha}(z) =\zeta \int_0^{z}\sqrt{3\Omega_\phi(z)\left(1+w_\phi(z)\right)}\frac{dz'}{1+z'}\,.
\end{equation}
The is assumes a canonical scalar field, but the argument can be repeated for phantom fields \cite{Phantom}, leading to 
\begin{equation} \label{eq:dalfa2}
\frac{\Delta\alpha}{\alpha}(z) =-\zeta \int_0^{z}\sqrt{3\Omega_\phi(z)\left|1+w_\phi(z)\right|}\frac{dz'}{1+z'}\,;
\end{equation}
the change of sign stems from the fact that one expects phantom filed to roll up the potential rather than down.

In the present work we'll focus on models with a constant equation of state $w_0$, and will constrain them using the following datasets
\begin{itemize}
\item Cosmological data: the Union2.1 Type Ia supernova dataset \cite{Union} and the compilation of Hubble parameter measurements from Farooq \& Ratra \cite{Farooq}.
\item Laboratory data: the atomic clock constraint on the current drift of $\alpha$ of Rosenband {\it et al.} \cite{Rosenband}, which we can write as
\begin{equation} \label{clocks}
\frac{1}{H_0}\frac{\dot\alpha}{\alpha} =(-2.2\pm3.2)\times10^{-7}\,.
\end{equation}
\item Astrophysical data: we will use both spectroscopic measurements of $\alpha$ of Webb {\it et al.} \cite{Webb} (a large dataset of 293 archival data measurements) and the smaller and more recent dataset of 11 dedicated measurements listed in Table \ref{table1}. The latter include the early results of the UVES Large Program for Testing Fundamental Physics \cite{LP1,LP3}, which is expected to be the one with a better control of possible systematics.
\end{itemize}

\begin{table}
\begin{center}
\begin{tabular}{|c|c|c|c|c|}
\hline
 Object & z & ${ \Delta\alpha}/{\alpha}$ & Spectrograph & Ref. \\ 
\hline\hline
3 sources & 1.08 & $4.3\pm3.4$ & HIRES & \protect\cite{Songaila} \\
\hline
HS1549$+$1919 & 1.14 & $-7.5\pm5.5$ & UVES/HIRES/HDS & \protect\cite{LP3} \\
\hline
HE0515$-$4414 & 1.15 & $-0.1\pm1.8$ & UVES & \protect\cite{alphaMolaro} \\
\hline
HE0515$-$4414 & 1.15 & $0.5\pm2.4$ & HARPS/UVES & \protect\cite{alphaChand} \\
\hline
HS1549$+$1919 & 1.34 & $-0.7\pm6.6$ & UVES/HIRES/HDS & \protect\cite{LP3} \\
\hline
HE0001$-$2340 & 1.58 & $-1.5\pm2.6$ &  UVES & \protect\cite{alphaAgafonova}\\
\hline
HE1104$-$1805A & 1.66 & $-4.7\pm5.3$ & HIRES & \protect\cite{Songaila} \\
\hline
HE2217$-$2818 & 1.69 & $1.3\pm2.6$ &  UVES & \protect\cite{LP1}\\
\hline
HS1946$+$7658 & 1.74 & $-7.9\pm6.2$ & HIRES & \protect\cite{Songaila} \\
\hline
HS1549$+$1919 & 1.80 & $-6.4\pm7.2$ & UVES/HIRES/HDS & \protect\cite{LP3} \\
\hline
Q1101$-$264 & 1.84 & $5.7\pm2.7$ &  UVES & \protect\cite{alphaMolaro}\\
\hline
\end{tabular}
\caption{\label{table1}Recent dedicated measurements of $\alpha$. Listed are, respectively, the object along each line of sight, the redshift of the measurement, the measurement itself (in parts per million), the spectrograph, and the original reference. The first measurement is the weighted average from 8 absorbers in the redshift range $0.73<z<1.53$ along the lines of sight of HE1104-1805A, HS1700+6416 and HS1946+7658, reported in \cite{Songaila} without the values for individual systems. The UVES, HARPS, HIRES and HDS spectrographs are respectively in the VLT, ESO 3.6m, Keck and Subaru telescopes.}
\end{center}
\end{table}

\section{Observational constraints}

We now use the above datasets to constrain the dynamical dark energy model with a constant equation of state $w_0$. The behaviour of $\alpha$ will be determined by Eq.(\ref{eq:dalfa}) for $w_0>-1$ and Eq.(\ref{eq:dalfa2}) for $w_0<-1$. Our main interest is in obtaining constraints on the $\zeta$--$w_0$ plane, and for this reason we will fix $H_0=70$ km/s/Mpc and $\Omega_{m0}=0.3$ (and assume a flat universe, so $\Omega_{\phi0}=0.7$). This choice of cosmological parameters is fully consistent with the supernova and Hubble parameter data we use. Moreover, we have explicitly verified that allowing $H_0$, $\Omega_m$ or the curvature parameter to vary (within observationally reasonable ranges) and marginalizing over these parameters does not significantly change our results: it is clear that the critical cosmological parameter here is $w_0$ itself.

\begin{figure}
\begin{center}
\includegraphics[width=3in]{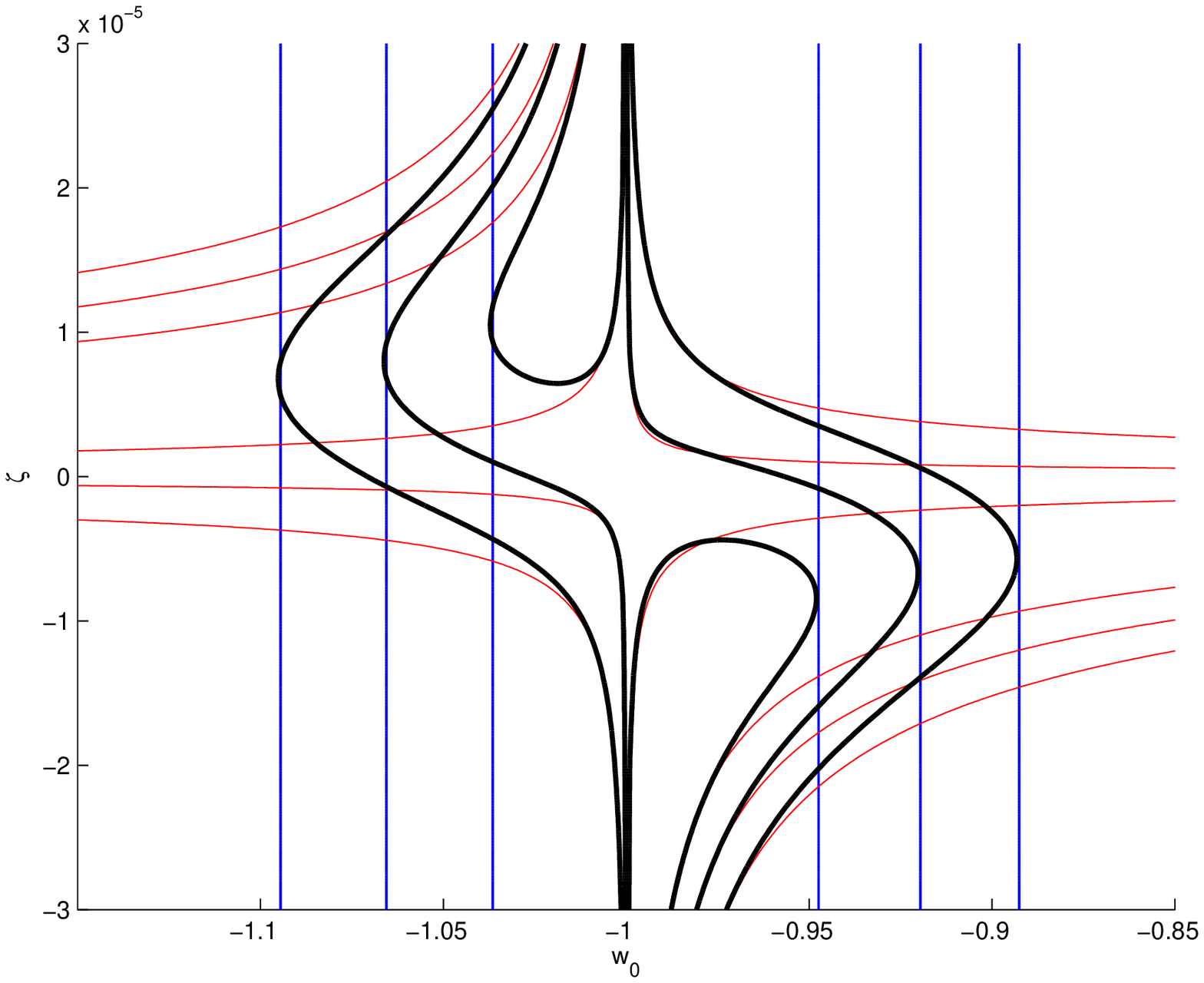}
\includegraphics[width=3in]{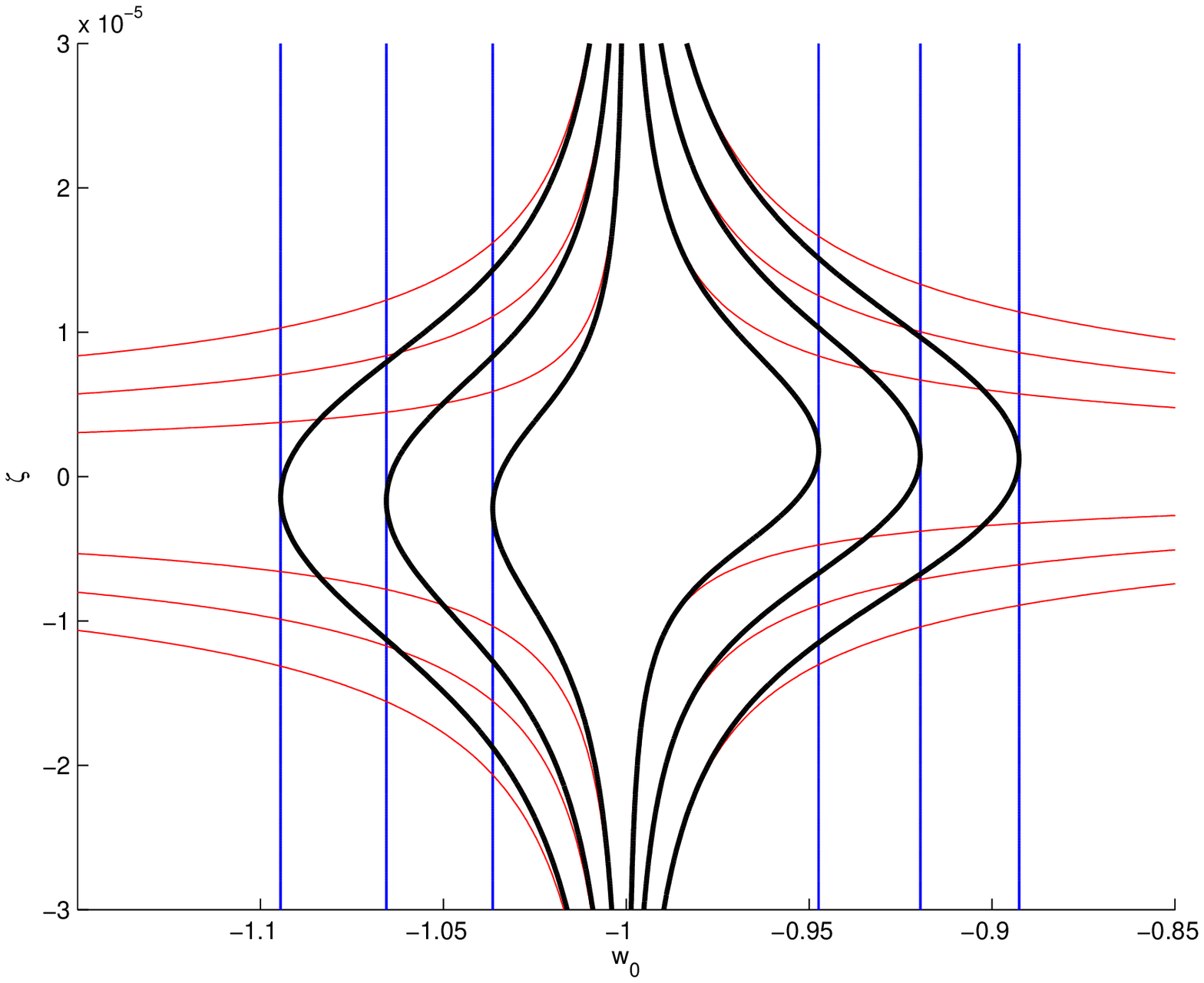}
\includegraphics[width=3in]{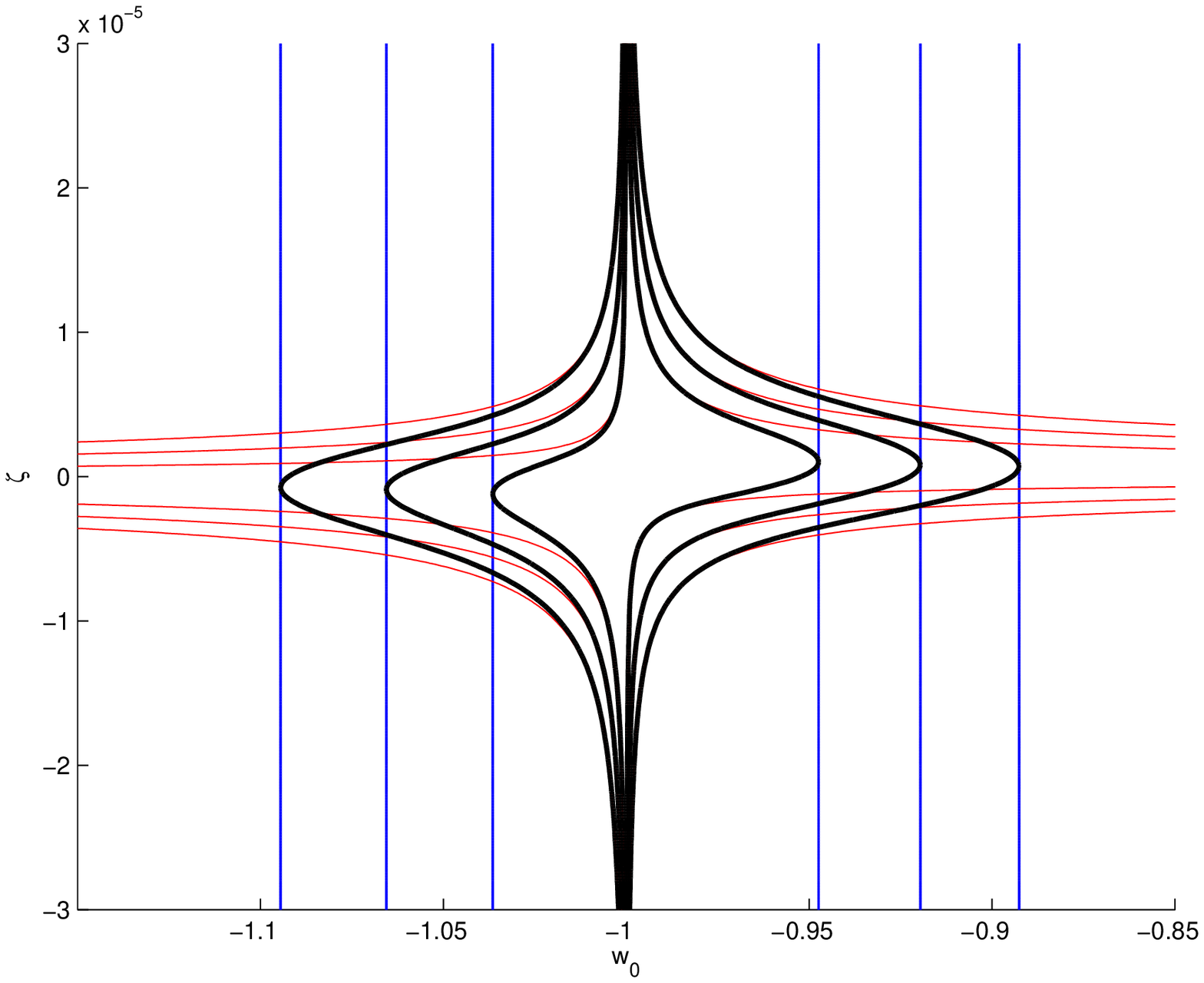}
\end{center}
\caption{\label{figalpha}One, two and three sigma constraints on the $\zeta-w_0$ plane from Webb {\it et al.} data (top panel), Table \protect\ref{table1} data (middle) and the atomic clock bound (bottom). In each panel the thin red lines correspond to the constraints from the astrophysical or clock data alone, the blue vertical ones correspond to the cosmological data (which constrain $w_0$ but are insensitive to $\zeta$) and the black thick lines correspond to the combined datasets.}
\end{figure}

We therefore consider a 2D grid of $\zeta$ and $w_0$ values, and use standard chi-square techniques to compare the models to the aforementioned datasets. Figure \ref{figalpha} shows the results of this comparison for the Webb {\it et al.} data (top panel), and for the Table \protect\ref{table1} data (middle panel)---in both cases, the constraints from the astrophysical data are shown by the thin red lines. The Webb data is not consistent with the null result \cite{Webb}, and we correspondingly find a one sigma preference for a non-zero coupling $\zeta$ (with a negative sign for a canonical field, or a positive sign for a phantom field). However, the data is compatible with the null result at two sigma. On the other hand, the Table \protect\ref{table1} data is fully compatible with the null result. We note that in the former case the reduced chi-square of the best-fit model is $\chi^2_{min,Webb}=1.04$, while in the latter case it is $\chi^2_{min,Table}=1.29$; this may be an indication that some of the uncertainties in the Table \protect\ref{table1} measurements are underestimated.

For comparison we also show in the bottom panel of Fig. \ref{figalpha}, in the same scale as before (and also in thin red lines), the local atomic clock constraint of  Rosenband {\it et al.} \cite{Rosenband}. For the models under consideration this translates into
\begin{equation} \label{clocks2}
\frac{1}{H_0}\frac{\dot\alpha}{\alpha} =\, \mp \, \zeta\sqrt{3\Omega_{\phi0}|1+w_0|}\,
\end{equation}
(with the $-$ and $+$ signs respectively corresponding to the canonical and phantom field cases), and it is clear from the plot that this is currently more constraining than the astrophysical measurements.

\begin{figure}
\begin{center}
\includegraphics[width=3in]{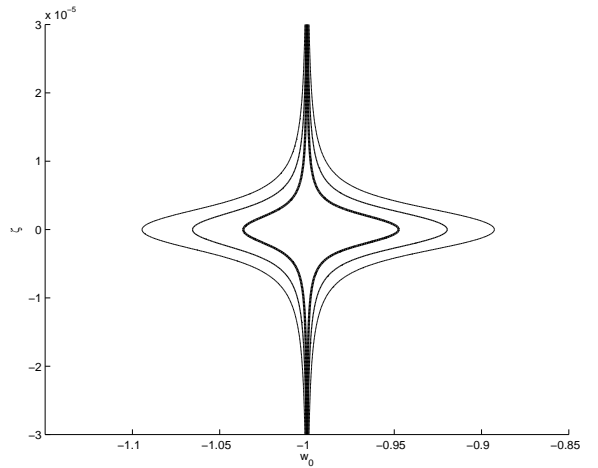}
\end{center}
\caption{\label{figfinal}One, two and three sigma constraints on the $\zeta-w_0$ plane from the full dataset considered in our analysis: Webb {\it et al.} data plus Table \protect\ref{table1} data plus atomic clock bound plus cosmological (Type Ia supernova and Hubble parameter) data. The reduced chi-square of the best fit is $\chi^2_{min,full}=0.97$.}
\end{figure}

The cosmological data we are considering is insensitive to $\zeta$. (Strictly speaking, a varying $\alpha$ does affect the luminosity of Type Ia supernovas, but as recently shown in \cite{Erminia2} for parts-per-million level $\alpha$ variations the effect is too small to have an impact on current datasets, and we therefore neglect it in the present analysis.) Naturally, the cosmological data does constrain $w_0$, effectively providing us with a prior on it. In Fig. \ref{figalpha} the cosmological data constraints are shown by the blue vertical lines, while the combined (cosmological plus astrophysical, or cosmological plus atomic clocks) constraints are shown by the thick black lines. Naturally, we can obtain tighter constraints by combining all datasets; this is straightforward to do since the Webb {\it et al.} and Table \ref{table1} measurements are independent. The results of this analysis are shown in Fig. \ref{figfinal}.

\begin{figure}
\begin{center}
\includegraphics[width=3in]{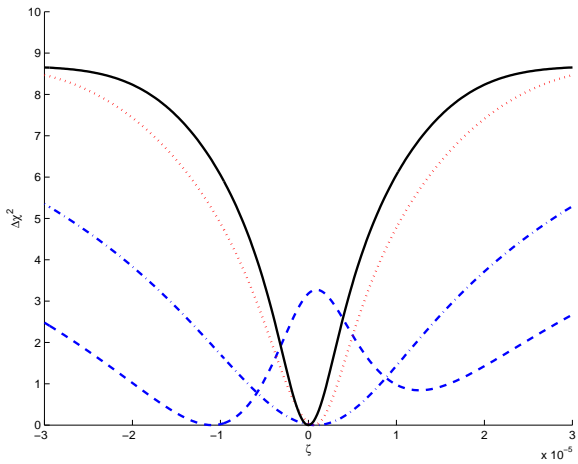}
\end{center}
\caption{\label{figmarg}1D likelihood for $\zeta$, marginalizing over $w_0$. Plotted is the value of $\Delta\chi^2=\chi^2-\chi^2_{min}$, for cosmological $+$ Webb data (blue dashed), cosmological $+$ Table \protect\ref{table1} data (blue dash-dotted), cosmological $+$ atomic clock data (red dotted) and the combination of all datasets (black solid).}
\end{figure}

Finally, in addition to constraints in the two-dimensional $\zeta$--$w_0$ plane it is also interesting to obtain 1D constraints on the coupling $\zeta$ by marginalizing over the dark energy equation of state $w_0$. The results of this analysis are shown in Fig. \ref{figmarg}. We confirm that in the case of the Webb {\it et al.} dataset there is a one-sigma preference for a non-zero coupling, while in the other cases the null result provides the best fit. Significantly, the combination of all datasets allows us to obtain a non-trivial constraint on $\zeta$. At the two-sigma ($95.4\%$) confidence level we find
\begin{equation} \label{newzetabound}
|\zeta|<5\times10^{-6}\,,
\end{equation}
significantly improving upon previous constraints. As previously mentioned, the atomic clock measurement of Rosenband {\it et al.} currently provides tighter constraints than the astrophysical measurements. This new bound is the main result of our analysis. (Nevertheless, we also note that at three-sigma $\zeta$ is unconstrained.) We can similarly obtain the 1D likelihood for $w_0$ by marginalizing over $\zeta$. In this case we find at the three-sigma ($99.7\%$) confidence level
\begin{equation} \label{neww0bound}
-1.05<w_0<-0.94\,,
\end{equation}
although this bound should be interpreted more cautiously given our assumptions on other cosmological parameters. We leave a more systematic exploration of the relevant parameter space for a subsequent publication.

\section{Conclusions and outlook}

We have used a combination of astrophysical spectroscopy and local laboratory tests of the stability of the fine-structure constant $\alpha$, complemented by background cosmological datesets, to constrain the simplest class of dynamical dark energy models where the same degree of freedom is responsible for both the dark energy and a variation of $\alpha$. In these models the redshift dependence of $\alpha$ depends both on a fundamental physics parameter (the dimensionless coupling $\zeta$ of the scalar field to the electromagnetic sector) and background 'dark cosmology' parameters---for the simplest class of models we studied these are the dimensionless dark energy density $\Omega_\phi$ and the dark energy equation of state $w_0$.

We obtained new, tighter constraints on the dimensionless coupling $\zeta$ of the scalar field to the electromagnetic sector. We note that these constraints are currently dominated by the atomic clock tests, which are only sensitive to the dark energy equation of state today. Thus a constant equation of state cosmological model is a reasonable assumption. Improvements in astrophysical measurements will allow more generic models to be constrained.

We have also pointed out how different currently available astrophysical measurements of $\alpha$ (specifically the archival data of Webb {\it et al.} and the dedicated measurements of Table \ref{table1}) lead to somewhat different constraints). This highlights the importance of obtaining improved astrophysical measurements of $\alpha$ (both in terms of statistical uncertainty and in terms of control over possible systematics), not only for their own sake but also because there can have a strong impact on dark energy studies. The next generation of high-resolution ultra-stable spectrographs such as ESPRESSO and ELT-HIRES will be ideal for this task. A roadmap for these studies is outlined in \cite{cjmGRG}, and more detailed forecasts of the future impact of these measurements may be found in \cite{Leite}.

\begin{acknowledgments}
This work was done in the context of project PTDC/FIS/111725/2009 (FCT, Portugal). CJM is also supported by an FCT Research Professorship, contract reference IF/00064/2012, funded by FCT/MCTES (Portugal) and POPH/FSE (EC).
\end{acknowledgments}

\bibliography{paper1}
\end{document}